\documentclass[conference]{IEEEtran}

\usepackage{tikz}
\usetikzlibrary{positioning, shapes.geometric, arrows}

\usepackage{cite}
\usepackage{amsmath,amssymb,amsfonts}
\usepackage{algorithmic}
\usepackage{graphicx}
\usepackage{textcomp}
\usepackage{xcolor}
\usepackage{tabularx}

\usepackage{url}

\usepackage{pifont}
\newcommand{\cmark}{\ding{51}}%
\newcommand{\xmark}{\ding{55}}%
\newcommand{\smark}{\ding{81}}%

\usepackage{tikz}
\usetikzlibrary{shapes.geometric, arrows}

\tikzstyle{startstop} = [rectangle, rounded corners, minimum width=3cm, minimum height=1cm,text centered, draw=black, fill=gray!30]
\tikzstyle{process} = [rectangle, minimum width=3cm, minimum height=1cm, text centered, draw=black, fill=blue!30]
\tikzstyle{decision} = [diamond, minimum width=3cm, minimum height=1cm, text centered, draw=black, fill=green!30]
\tikzstyle{arrow} = [thick,->,>=stealth]

\usepackage{tikz}
\usetikzlibrary{shapes.geometric, arrows}

\tikzstyle{startstop} = [rectangle, rounded corners, minimum width=3cm, minimum height=1cm,text centered, draw=black, fill=gray!30]
\tikzstyle{process} = [rectangle, minimum width=3cm, minimum height=1cm, text centered, draw=black, fill=blue!30]
\tikzstyle{cloud} = [rectangle, minimum width=3cm, minimum height=1cm, text centered, draw=black, fill=orange!30]
\tikzstyle{arrow} = [thick,->,>=stealth]

\begin{document}


\title{Generative AI in Multimodal User Interfaces: Trends, Challenges, and Cross-Platform Adaptability}

\author{
    \IEEEauthorblockN{Jan Bieniek\IEEEauthorrefmark{1}, Mohamed Rahouti\IEEEauthorrefmark{2}, Dinesh C. Verma\IEEEauthorrefmark{3}}
    \IEEEauthorblockA{\IEEEauthorrefmark{1}Department of Computer Science, Fordham University, New York, NY 10023, USA \\
    Email: jbieniek@fordham.edu}
    \IEEEauthorblockA{\IEEEauthorrefmark{2}Department of Computer Science, Fordham University, New York, NY 10023, USA \\
    Email: mrahouti@fordham.edu}
    \IEEEauthorblockA{\IEEEauthorrefmark{2}IBM TJ Watson Research Center, Yorktown Heights, NY 10598, USA \\
    Email: dverma@us.ibm.com}
}

\maketitle

\begin{abstract}
As the boundaries of human-computer interaction expand, Generative AI emerges as a key driver in reshaping user interfaces (UIs), introducing new possibilities for personalized, multimodal, and cross-platform interactions. This integration reflects a growing demand for more adaptive and intuitive UIs that can accommodate diverse input types—text, voice, video—and deliver seamless experiences across devices. This paper explores the integration of Generative AI in modern UIs, examining historical developments and focusing on multimodal interaction, cross-platform adaptability, and dynamic personalization. A central theme is the ``interface dilemma," which addresses the challenge of designing effective interactions for multimodal large language models (LLMs), assessing the trade-offs between graphical, voice-based, and immersive interfaces. The paper further evaluates lightweight frameworks tailored for mobile platforms, spotlighting the role of mobile hardware in enabling scalable, multimodal AI. Technical and ethical challenges, including context retention, privacy concerns, and balancing cloud and on-device processing, are thoroughly examined. Finally, the paper outlines future directions, such as emotionally adaptive interfaces, predictive AI-driven UIs, and real-time collaborative systems, underscoring Generative AI’s potential to redefine adaptive, user-centric interfaces across platforms.
\end{abstract}



\section{Introduction}

User interfaces have significantly evolved over the last few decades. From early text-based systems to modern graphical and multimodal interfaces, the developments in human-computer-driven interactions were strongly driven by advancements in software and hardware. This review examines how Generative AI, particularly multimodal large language models (LLMs) \cite{brown2020language}, is poised to further transform UI design by enabling dynamic personalization, context retention, and efficient scalability. It seems inevitable that with the advent of highly capable LLMs being accessible on every device, the way people interact with technology is going to adapt. This raises a few critical questions: What is the ideal interface for users to interact with AI-powered systems? Is there going to be a single trend or will interfaces adapt to each type of application specifically? How will the availability of immersive technologies like Virtual Reality (VR) glasses influence this? How long can it take people to adapt to a completely new interface experience? Some of the biggest technology companies have been experiencing with introducing new ways to interact with technology, but in the end, they converge to very similar styles and ideas.

We will also address the limitations of current frameworks and the challenges of implementing AI in this domain, particularly when constrained by hardware like mobile phones.

\subsection{Problem Statement: The Interface Dilemma}

Since the release of Chat GPT by Open AI, there has been a massive spike in interest in AI-driven applications. The chatbot-like interface, which was popularized by Chat GPT has quickly become the standard for human-AI interactions. Despite recent developments of multimodal LLMs, chat-based interactions are still the most popular ones, despite their limitations. As a result, most applications are limited to a very similar user interface and require high-quality user input to help the LLM understand the context or mood. One example of such applications can be voice assistants such as Alexa, Google Assistant, or Siri. Apple first integrated Siri into its ecosystem with the iPhone 4S in 2011. Since then, both the hardware and software it runs, as well as its capabilities, have improved massively. However, the way users interact with Siri has stayed exactly the same. Similarly, Google Assistant and Alexa, despite being competing products, offer exactly almost identical experiences when it comes to interacting with their device.

We already possess powerful multimodal LLMs capable of processing text, images, and voice. However, the challenge lies in determining the best interface for human-computer interaction. Should it be console-based, GUI-driven, or even integrated into VR glasses? The core issue revolves around formulating an interface that is intuitive and leverages the full capabilities of AI-driven multimodal systems.

\subsection{Objectives}

This review article aims to critically analyze and synthesize the state-of-the-art advancements in AI-driven user interface (UI) design, with a particular emphasis on optimizing interfaces for multimodal LLMs. The central focus is to explore how various forms of interaction, including text, voice, and video, can be seamlessly integrated into UIs to enhance human-computer interaction. Additionally, this article investigates the application of lightweight frameworks, particularly in the context of mobile devices, and assesses the potential of mobile phone hardware (voice, video, text) as a primary platform for scalable, efficient, and user-friendly AI interfaces.

\begin{table}[h!]
\centering
\caption{List of key acronyms used in the paper.}
\begin{tabular}{|c|l|}
\hline
\textbf{Acronym} & \textbf{Definition} \\ \hline
AI & Artificial Intelligence \\ \hline
AR & Augmented Reality \\ \hline
BCI & Brain-Computer Interface \\ \hline
CLI & Command-Line Interface \\ \hline
GUI & Graphical User Interface \\ \hline
IoT & Internet of Things \\ \hline
LLM & Large Language Model \\ \hline
NLP & Natural Language Processing \\ \hline
NPU & Neural Processing Unit \\ \hline
UI & User Interface \\ \hline
VR & Virtual Reality \\ \hline
\end{tabular}
\label{tab:acronyms}
\end{table}

\begin{table*}[h!]
\centering
\caption{Comparison of review articles about Generative AI in multimodal user interfaces. \cmark, \xmark, and \smark indicate that the topic is well-covered, uncovered, and partially covered, respectively.} 
\label{tab:review_comparison}
\footnotesize
\begin{tabular}{|p{2.2cm}|p{0.6cm}|p{5.8cm}|p{1.5cm}|p{1.3cm}|p{1.6cm}|p{1.5cm}|}
\hline
\textbf{Reference} 
& \textbf{Year} 
& \textbf{Focus} 
& \textbf{Multimodal Interaction}
& \textbf{Generative AI}
& \textbf{Cross-Platform}
& \textbf{Challenges \& Trends}
\\ \hline
\hline
Bieniek \textit{et al.} (This Paper) & 2024 & A review of integrating Generative AI in modern computer UIs, focusing on multimodal interaction, cross-platform compatibility, and dynamic personalization & \cmark  &  \cmark &  \cmark & \cmark \\ \hline
Jones \textit{et al.} \cite{jones2024multimodal} & 2024 & Examines how multimodal LLMs ground language in comparison to human interactions & \cmark & \xmark & \xmark & \smark \\ \hline
Munikoti \textit{et al.} \cite{munikoti2024generalist} & 2024 & Reviews architectures, challenges, and opportunities in multimodal AI, including transformer-based architectures for fusion & \cmark & \cmark & \smark & \cmark \\ \hline
Huang \textit{et al.} \cite{huang2024unlocking} & 2024 & Focuses on adaptive user experience with Generative AI and dynamic personalization & \xmark & \cmark & \cmark & \smark \\ \hline
Kim \textit{et al.} \cite{kim2021multimodal} & 2021 & Reviews multimodal interaction systems based on IoT and augmented reality & \cmark & \xmark & \cmark & \cmark \\ \hline
Lu \textit{et al.} \cite{lu2024ai} & 2024 & Reviews AI in user experience design, with an emphasis on human-centered AI applications & \xmark & \cmark & \cmark & \xmark \\ \hline
Pyarelal \textit{et al.} \cite{pyarelal2018automating} & 2018 & Discusses automating the design of UIs using AI technologies & \smark & \cmark & \smark & \smark \\ \hline
Su \textit{et al.} \cite{su2023recent} & 2023 & Reviews advancements in multimodal human-robot interaction, with emphasis on combining voice, gestures, and facial recognition & \cmark & \xmark & \xmark & \cmark \\ \hline
Zhang \textit{et al.} \cite{zhang2024unveiling} & 2024 & Evaluates the impact of multimodal interactions on user engagement in AI-driven conversations & \cmark & \cmark & \xmark & \cmark \\ \hline
Bandi \textit{et al.} \cite{bandi2023power} & 2023 & Reviews generative AI models, input-output formats, evaluation metrics, and technical challenges & \xmark & \cmark & \smark & \cmark \\ \hline
\end{tabular}
\end{table*}

\subsection{Scope}

This article focuses on the intersection of AI and UI design, with particular attention to multimodal interaction with LLMs. The review encompasses the technological frameworks that enable AI-driven UI optimization, especially for mobile platforms, and evaluates lightweight approaches for integrating voice, video, and text modalities. The scope is limited to UIs that support multimodal interaction and dynamic personalization, excluding non-interactive or single-modal systems. Furthermore, the review assesses the feasibility of deploying such systems on mobile hardware, considering processing limitations, scalability challenges, and the need for efficient resource management.

\subsection{Related Papers}


Table \ref{tab:review_comparison} provides a structured comparison of various review articles on generative AI, multimodal interfaces, and cross-platform adaptability. Each reference contributes uniquely to the body of research, highlighting different aspects of AI-driven user interfaces.

For instance, Jones et al. \cite{jones2024multimodal} focus on the comparison between multimodal LLMs and human language grounding, providing valuable insights into how AI systems perceive and interpret multimodal inputs. However, their work lacks a detailed discussion on cross-platform adaptability and the role of generative AI in modern interfaces. In contrast, Munikoti et al. \cite{munikoti2024generalist} delve deep into the architecture of multimodal AI systems, such as transformer-based models, and discuss their potential across different platforms. While their review thoroughly covers multimodal interaction, it only partially addresses the cross-platform challenges.

Huang et al. \cite{huang2024unlocking} contribute by emphasizing adaptive user experiences through generative AI. Their review offers insights into how dynamic personalization can be achieved using generative AI but lacks an extensive exploration of multimodal systems. Similarly, Kim et al. (2021) focus on multimodal interaction systems in Internet of Things (IoT) and Augmented Reality (AR), emphasizing cross-platform challenges and offering a robust view of system interoperability but without a focus on generative AI \cite{kim2021multimodal}. On the other hand, Lu et al. \cite{lu2024ai} concentrate on human-centered AI in user experience design, which is highly relevant to AI-driven interfaces but does not deeply explore multimodal interaction.

While some works, such as Su et al. \cite{su2023recent}, offer a comprehensive review of multimodal human-robot interaction, they miss out on the cross-platform adaptability required in broader AI-driven systems. Bandi et al. \cite{bandi2023power}, though thorough in covering generative AI models and technical challenges, only partially address cross-platform issues and multimodal interaction.

While these articles contribute significantly to the fields of generative AI and multimodal user interfaces, my attached paper offers a more comprehensive review by covering multimodal interaction, generative AI integration, cross-platform adaptability, and the emerging trends and challenges in AI-driven UIs. This distinguishes the paper as a broader, more inclusive review, bridging gaps not fully addressed by other works.

\subsection{Paper Structure}

The structure of this paper is as follows: Section II covers the interface dilemma, highlighting challenges in designing intuitive multimodal LLM interfaces. Section III reviews the evolution of user interfaces, noting limitations of current AI-driven platforms. Section IV discusses application frameworks and AI integration, focusing on Generative AI's role in cross-platform adaptability and mobile integration. Section V examines the development of new multimodal UIs enabled by Generative AI, addressing privacy, mobile processing efficiency, and context retention challenges. Section VI explores limitations and challenges in AI-driven interfaces and outlines future trends. Section VII identifies evaluation metrics for AI-driven multimodal UIs and Section VIII concludes the paper. Lastly, the list of key acronyms used in this paper is given in Table \ref{tab:acronyms}.

\section{Problem Statement: The Interface Dilemma}

The release of ChatGPT by OpenAI marked a pivotal moment in AI-driven applications, popularizing a chatbot-like interface as the standard for human-AI interactions. ChatGPT's rapid adoption demonstrated the appeal of natural language interaction, allowing users to communicate with AI through a simple, conversational UI. However, this approach, while intuitive and effective for text-based communication, exposes limitations in scalability and adaptability, particularly when applied to more complex, multimodal LLMs \cite{li2024map}. Despite the rapid advancements in AI capabilities, the chatbot interface remains largely constrained by its dependence on high-quality user input to understand and process context, mood, and intent accurately.

\subsection{Limitations of Chat-Based Interfaces}

Chat-based interfaces, as exemplified by systems like ChatGPT, remain the most widely used form of human-AI interaction. However, this paradigm suffers from significant limitations when applied to more sophisticated multimodal LLMs, which are designed to process not only text but also images, video, and audio. The chatbot model is inherently linear, lacking the flexibility to seamlessly integrate multiple input types. As a result, users must manually structure their inputs to ensure the LLM can interpret and generate relevant outputs across modalities \cite{groner2024investigating}. Furthermore, the dependency on text-based input places a burden on users to provide detailed, well-structured prompts, reducing accessibility for individuals less familiar with technical language or who prefer more dynamic interaction modes.

\subsection{Emergence of Voice-Based Interaction}

Alongside chat-based interfaces, voice-driven systems have become increasingly popular, driven by voice assistants such as Siri, Alexa, and Google Assistant. These systems allow users to interact with AI using natural speech, making them more accessible and user-friendly for a wider audience. The introduction of live conversations with LLMs expands this functionality further, allowing for real-time, multimodal interaction where voice inputs are integrated into the processing pipeline of advanced AI models \cite{liu2024mental}. However, despite the growing prominence of voice-based interfaces, they face challenges in understanding context, mood, and complex queries, particularly when compared to human interactions \cite{jones2024multimodal}. For example, while voice assistants can handle basic commands, they often struggle with maintaining long-term context or interpreting nuanced user inputs in real-time \cite{mridha2021brain}.

Recent efforts to enhance live conversations with LLMs involve incorporating context retention mechanisms and emotion detection algorithms to improve the accuracy and relevance of responses. This is particularly evident in the development of systems that combine voice, text, and visual inputs, allowing the AI to dynamically adjust to the user’s changing needs. However, achieving fluidity in these interactions remains a challenge, as systems must balance the cognitive load on users with the AI's capacity to process and respond to multimodal inputs in real time \cite{fallman2003design}.

\subsection{The Interface Dilemma: Multimodal LLMs and Interaction Modes}

Multimodal LLMs have unlocked unprecedented capabilities in processing diverse data types—text, images, audio, and video—but their integration into user-friendly interfaces remains a complex challenge. The key question is: what is the ideal interface for human-computer interaction with these powerful AI systems? Current approaches range from console-based environments, which are highly flexible but require technical expertise, to Graphical User Interfaces (GUIs), which offer accessibility but may not fully leverage the multimodal processing abilities of LLMs \cite{moore2023empowering}. Additionally, emerging technologies such as VR and AR present novel opportunities for immersive, multimodal interaction but require sophisticated hardware and interfaces that are not yet fully mainstream \cite{sharma1998toward, huang2024large}. Each modality offers distinct advantages and faces unique challenges.

\begin{table*}[h]
\centering
\caption{Comparison of multimodal LLM interaction methods.}
\begin{tabular}{|c|c|c|c|c|c|}
\hline
\textbf{Interaction Mode} & \textbf{User Accessibility} & \textbf{Input Complexity} & \textbf{Response Accuracy} & \textbf{System Requirements} & \textbf{Scalability} \\ \hline
\textbf{Text-based} & High & Low & Moderate & Low & High \\ \hline
\textbf{Voice-based} & Moderate & Moderate & High & Moderate & High \\ \hline
\textbf{Video-based} & Low & High & High & High & Low \\ \hline
\textbf{Immersive (VR/AR)} & Low & High & High & Very High & Low \\ \hline
\end{tabular}
\label{tab:multimodal_comparison}
\end{table*}

As shown in Table \ref{tab:multimodal_comparison}, the comparison of multimodal LLM interaction methods highlights significant differences in user accessibility, input complexity, response accuracy, system requirements, and scalability. Text-based interaction offers high accessibility with low input complexity, but its response accuracy and scalability are moderate. Voice-based systems, while providing more dynamic interaction, face moderate input complexity and system requirements, though they perform well in scalability and accuracy. Video-based interaction introduces higher complexity and system demands, resulting in reduced scalability, despite achieving high accuracy. Finally, immersive VR/AR technologies, although promising in delivering a highly interactive experience with accurate responses, are constrained by very high system requirements and low scalability, making them less accessible for widespread use.

\subsubsection{Console-based interfaces} 
Traditionally favored by developers and power users, console-based interfaces offer flexibility and control, particularly in environments where precise input is required. For multimodal LLMs, consoles allow for direct command execution and the integration of various input forms. However, their complexity makes them inaccessible to general users, limiting their applicability in broader consumer applications \cite{torricelli2024role}.

\subsubsection{Graphical User Interfaces (GUIs)} 
GUIs are the most common form of interaction for non-technical users, allowing for intuitive, point-and-click navigation. However, the limitations of GUIs become evident when handling complex multimodal interactions. For example, GUIs designed for multimodal LLMs often struggle to integrate seamless transitions between text, image, and voice inputs. Additionally, the GUI model often lacks context retention, making it difficult for the system to remember previous interactions or personalize future responses \cite{li2024user, huang2024large}.

\subsubsection{Immersive technologies (VR/AR)} 
Immersive interfaces, such as VR and AR, offer potential solutions to the multimodal dilemma by creating interactive, 3D environments where users can engage with AI systems more naturally. In VR or AR settings, users can interact with objects and receive multimodal feedback (visual, auditory, haptic) from the LLM. However, these interfaces require expensive hardware and pose usability challenges, particularly for mobile applications, where the computational load of running real-time multimodal AI on limited hardware is significant \cite{groner2024investigating}.

\subsection{Designing Intuitive Multimodal Interfaces}

The core of the interface dilemma lies in designing an intuitive, user-friendly interface that can fully leverage the multimodal capabilities of LLMs. This involves determining not only the optimal interaction modality (text, voice, video) but also how to seamlessly integrate these modalities into a single, coherent user experience \cite{wang2024comprehensive}. Current research suggests that a hybrid approach—combining the simplicity of GUIs with the versatility of multimodal inputs—may provide the most practical solution \cite{torricelli2024role}. In such systems, users could begin an interaction with a text prompt, seamlessly transition to voice or video input, and receive contextually relevant responses from the LLM without requiring separate interfaces for each modality.

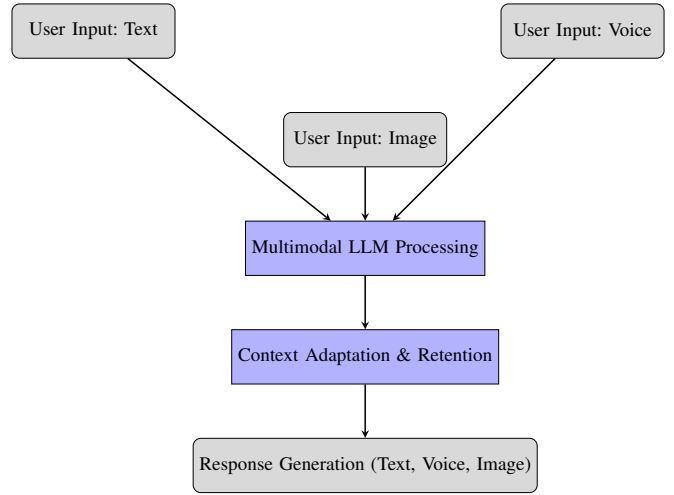
\begin{figure}[h]
\centering
\resizebox{\columnwidth}{!}{
\begin{tikzpicture}[node distance=2cm]

\node (text) [startstop] {User Input: Text};
\node (voice) [startstop, right of=text, xshift=7cm] {User Input: Voice};
\node (image) [startstop, below of=text, xshift=5cm] {User Input: Image};
\node (preprocess) [process, below of=image] {Multimodal LLM Processing};
\node (adapt) [process, below of=preprocess] {Context Adaptation \& Retention};
\node (response) [startstop, below of=adapt] {Response Generation (Text, Voice, Image)};

\draw [arrow] (text) -- (preprocess);
\draw [arrow] (voice) -- (preprocess);
\draw [arrow] (image) -- (preprocess);
\draw [arrow] (preprocess) -- (adapt);
\draw [arrow] (adapt) -- (response);

\end{tikzpicture}
}
\caption{Graphical representation of hybrid interface model.}
\label{fig:hybrid_interface}
\end{figure}

Figure \ref{fig:hybrid_interface} demonstrates the hybrid interface model, where users can interact with the system via multiple input types, including text, voice, and image. The inputs are processed through a multimodal LLM, followed by context adaptation and retention to ensure continuity in the interaction. The system then generates an appropriate response, which can be delivered in the form of text, voice, or image, depending on the input modality. This model highlights the flexibility of hybrid interfaces in dynamically handling various user inputs to enhance the overall interaction experience.

Moreover, advances in context retention algorithms are making it possible for multimodal LLMs to remember past interactions and adjust their responses based on the user’s behavior over time \cite{sarkar2023will}. These advancements are crucial for improving personalization, enhancing user satisfaction, and reducing the cognitive load associated with multimodal interactions.

Overall, the interface dilemma is a multifaceted problem that highlights the growing gap between the capabilities of multimodal LLMs and the limitations of current UIs. While chat-based and voice-driven interfaces dominate the landscape, they are ill-suited to fully harness the potential of multimodal systems. The challenge moving forward lies in designing versatile, intuitive interfaces that can dynamically handle text, voice, and visual inputs, while maintaining context and providing a personalized user experience. The development of such interfaces will be critical for the next generation of AI-driven applications, particularly in fields such as education, healthcare, and entertainment \cite{moore2023empowering, groner2024investigating}.

\section{History and Evolution of User Interfaces}

UIs have undergone significant transformations over the decades, evolving from basic text-based input systems to sophisticated multimodal platforms. This evolution reflects both advances in computing power and changing user expectations. This section provides an in-depth look at the historical progression of UIs, the current state of modern interfaces, and their limitations in the context of AI-driven systems, particularly multimodal LLMs.

\subsection{Early Interfaces}

In the earliest stages of human-computer interaction, UIs were primarily text-based and relied on manual command-line input. Users had to type specific commands to interact with computers, making the learning curve steep and restricting accessibility to a technically adept audience. Examples include the Command-Line Interface (CLI), where users had direct but complex control over the system's functions through text commands. Notable systems such as UNIX and MS-DOS embodied this interface model \cite{hartson1989human}.

The development of GUIs in the 1970s and 1980s marked a revolutionary step in UI design. GUIs offered a more intuitive approach, replacing text commands with visual representations, such as icons, windows, and menus. Xerox PARC is often credited with pioneering the first GUI, which later influenced the design of both the Apple Macintosh and Microsoft Windows operating systems \cite{chin1988development}. GUIs democratized computing, making it accessible to a broader audience.

\begin{table*}[h]
\centering
\caption{Evolution of user interfaces over time.}
\begin{tabularx}{\textwidth}{|c|c|X|X|c|}
\hline
\textbf{Era} & \textbf{UI Type} & \textbf{Key Innovations} & \textbf{Interaction Modalities} & \textbf{Example Systems} \\ \hline
\textbf{1960s-1970s} & Text-based (CLI) & CLI, keyboard inputs & Text & UNIX, MS-DOS \\ \hline
\textbf{1980s-1990s} & Graphical (GUI) & Point-and-click, windows, icons & Mouse, Text & Xerox PARC, Windows 95 \\ \hline
\textbf{2000s} & Touch-based & Touchscreens, gesture controls & Touch, Text & iPhone, Android \\ \hline
\textbf{2010s} & Voice-based & Voice recognition, NLP & Voice, Text & Siri, Alexa \\ \hline
\textbf{2020s} & Multimodal (LLMs) & AI-driven personalization, multimodal input (text, voice, image) & Text, Voice, Video & ChatGPT, Google Assistant \\ \hline
\end{tabularx}
\label{tab:evolution_ui}
\end{table*}

\subsection{Modern Interfaces}

The evolution of UIs continued with the rise of modern interfaces, which now support a range of input methods beyond text and graphical elements. Today's UIs incorporate touchscreens, voice recognition, and gesture-based inputs, reflecting the ongoing shift toward more natural and seamless forms of interaction \cite{mridha2021brain, sharma1998toward}. These advancements are exemplified by smartphones and devices like Amazon’s Alexa, which allow users to interact via voice commands, and tablets and smartphones, which rely on touch-based interactions.

With the advent of multimodal LLMs, current interfaces must handle inputs from multiple modalities, such as text, voice, and images, while dynamically adapting to user needs. This has brought about significant challenges, particularly in determining the most suitable interface for these AI-driven systems. For example, should users interact with LLMs through a console-based interface, a traditional GUI, or an immersive technology like VR or AR? Each modality offers unique benefits and limitations \cite{fallman2003design, moore2023empowering}.

As shown in Table \ref{tab:evolution_ui}, user interfaces have evolved significantly over time, driven by technological advancements. In the 1960s-1970s, text-based interfaces like the CLI dominated, relying on keyboard inputs. The 1980s-1990s saw the rise of GUI, introducing point-and-click functionality with systems like Windows 95. In the 2000s, touch-based interfaces became prevalent with devices like the iPhone, offering more intuitive interaction through touchscreens and gestures. By the 2010s, voice-based interfaces, enabled by natural language processing (NLP), gained traction with systems like Siri and Alexa. The 2020s introduced multimodal interfaces, combining text, voice, and video for richer interaction, exemplified by platforms like ChatGPT and Google Assistant.

\subsection{Interface Limitations}

Despite the progress in UI design, several limitations persist. One major drawback of many modern interfaces is their inability to retain user context across sessions. For instance, most current UIs are session-based, meaning they do not remember user preferences or interaction history, which limits their ability to offer personalized experiences \cite{groner2024investigating}. This is particularly challenging in the context of AI-driven systems, where personalization is key to enhancing user satisfaction and system efficiency.

Additionally, the emergence of multimodal LLMs raises critical questions about the future of interface design. The decision on whether to prioritize console-based, GUI-driven, or immersive technologies remains unresolved. Each modality presents trade-offs in terms of usability, accessibility, and processing requirements. For example, console-based interfaces may offer flexibility for developers but are less accessible to non-technical users, while GUIs provide visual simplicity but may struggle to handle complex multimodal inputs efficiently \cite{li2024map}.

Moreover, immersive technologies like VR and AR have the potential to offer richer, more interactive experiences, but they come with hardware and usability challenges that make them impractical for widespread use in everyday applications, especially those deployed on mobile devices \cite{li2024user}.

\subsection{Challenges for Multimodal LLMs}

Multimodal LLMs introduce a new layer of complexity to UI design. Unlike traditional interfaces that handle one form of input (e.g., text or touch), multimodal systems must seamlessly integrate multiple inputs—such as voice, text, and video—while maintaining contextual coherence. For instance, a user might begin an interaction with voice commands but switch to text inputs mid-session. Ensuring that the system correctly interprets and responds to these mixed inputs remains a significant technical challenge \cite{sarkar2023will, torricelli2024role}.

In addition, the resource constraints of mobile hardware further complicate the deployment of multimodal UIs. Processing multimodal inputs requires significant computational power, which can strain mobile devices with limited processing capabilities. This necessitates the development of lightweight frameworks that can efficiently manage multimodal interactions without sacrificing performance or user experience \cite{li2024user}.

\section{Current App Frameworks and AI Integration}

As AI continues to transform the digital landscape, current application frameworks are evolving to accommodate more intelligent, adaptable interfaces. The integration of Generative AI with existing app frameworks offers unprecedented opportunities to enhance user experience through dynamic interaction modes, while also presenting unique technical and operational challenges. This section examines the tech stacks commonly used in modern UIs and explores how Generative AI integration redefines the boundaries of cross-platform functionality and real-time interaction.

\subsection{Tech Stack Overview}


The current technology stack for UIs integrates server-client communication models with cross-platform frameworks, enabling applications to function seamlessly across devices like mobile phones, tablets, and desktop computers \cite{hartson1989human}. Cross-platform development frameworks, such as React Native and Flutter, offer consistent interfaces and enhanced scalability by reducing development effort across various operating systems. Generative AI introduces new dynamics to these stacks by incorporating modalities like voice, text, and image, which must be synchronized into a cohesive interface. Integrating Generative AI with these UIs can enable real-time adaptation, transforming user interactions by adding layers of personalization, engagement, and responsiveness \cite{Wang2024}. This setup, however, requires careful consideration of server load balancing, device compatibility, and efficient data processing to maintain a seamless user experience, especially on resource-limited mobile devices \cite{Moore2023}.

Integrating Generative AI into the existing UI tech stack further demands a robust backend infrastructure to support AI model inference and processing capabilities, especially for real-time interactions. Typically, these integrations rely on cloud-based AI services, such as AWS SageMaker, Google AI Platform, or Microsoft Azure Machine Learning, which offer scalable solutions to host and manage complex AI models \cite{li2024user}. However, balancing between on-device and cloud processing is crucial to ensure optimal performance, particularly for latency-sensitive applications. Mobile and IoT devices benefit from edge computing solutions that enable localized, efficient data handling to reduce dependency on remote servers, minimizing delay in data processing and enhancing privacy through localized data storage \cite{torricelli2024role}. This hybrid model of cloud-edge collaboration not only optimizes the tech stack for multimodal AI processing but also extends device capabilities, providing users with real-time, personalized experiences without compromising on speed or responsiveness.

\subsection{Generative AI for Personalization}

Generative AI models, like LLMs, enable real-time adaptation by generating content tailored to user behaviors and preferences. Through analysis of user inputs—whether in text, voice, or image format—these models can dynamically adjust the interface or content, offering users a personalized experience that evolves based on their actions \cite{kim2021multimodal}. Mobile devices, given their ubiquitous presence and support for multimodal capabilities, serve as an ideal platform to test lightweight AI-driven interfaces. These devices can utilize AI models to recognize voice commands, analyze visual inputs, and respond in natural language, all within a compact form factor. The adaptability of Generative AI on mobile platforms has shown potential in applications ranging from personalized recommendations in media to adaptive learning in educational apps, allowing for scalable AI deployment on devices with limited processing power \cite{huang2024unlocking}.

The effectiveness of Generative AI for personalization relies heavily on its capacity to retain contextual information across interactions, which enables a deeper level of user-specific customization. Huang \cite{huang2024generating} contributes to this personalization by demonstrating how AI assistants can generate user experiences tailored to diverse personas, allowing designers to address varied user needs and behaviors more effectively. This persona-based approach to AI-driven interaction highlights the importance of customized personas in creating engaging and relevant user experiences. Building on the ability of AI to create persona-based interactions, advanced Generative AI models further expand personalization by remembering user preferences, interaction history, and adapting to nuanced behavioral patterns over time.

For instance, in e-commerce applications, Generative AI can provide tailored product suggestions based on previous purchases, browsing habits, and current trends \cite{Su2023}. Furthermore, as user behavior evolves, these models are capable of dynamically updating their recommendations or adjusting the user interface to better suit changing preferences, all while operating within the constraints of mobile processing power \cite{zhang2024unveiling}. By utilizing edge computing and federated learning approaches, these systems can execute personalization tasks directly on the device, thus optimizing response times and ensuring privacy by keeping user data local \cite{Bandi2023}. This approach not only improves the efficiency and responsiveness of personalized AI interactions but also aligns with increasing demands for data security and privacy in user-centered applications.

\subsection{The Function Matching Problem}

A significant challenge in AI-driven UIs lies in matching generated content with appropriate application services, a problem exacerbated in multimodal systems where alignment between inputs (e.g., text, voice, and video) and functional outputs is essential. This function matching problem is particularly pronounced in multimodal interfaces, where each input type (e.g., voice command or video analysis) must accurately align with its corresponding output function within the system \cite{zhang2024unveiling}. Ensuring the generated content or commands map effectively to the application's services requires robust context management, real-time adaptability, and accurate interpretation of each modality \cite{Bandi2023}. For example, a user’s spoken command might need to trigger specific visual responses or textual feedback, depending on the interaction history and current context. This issue presents a substantial hurdle in designing cohesive multimodal UIs, where efficient synchronization across modalities remains crucial to provide a seamless, intuitive user experience \cite{groner2024investigating}.

The function matching problem also involves addressing inconsistencies in AI-generated outputs, which may arise due to variations in user intent, language ambiguities, or differences in input modalities. For instance, when a user’s voice command ambiguously references an action that could apply to multiple functionalities (e.g., ``open" for both files and applications), the system must employ advanced disambiguation techniques, such as context-aware NLP and probabilistic reasoning, to interpret the correct function \cite{torricelli2024role}. Additionally, for applications requiring real-time multimodal coordination—such as those in AR and VR environments—the challenge of function matching intensifies as the system must process diverse, simultaneous inputs while ensuring synchronous outputs \cite{Wang2024}. Leveraging reinforcement learning and adaptive feedback loops, systems can refine their accuracy over time, gradually improving the alignment between multimodal inputs and the associated functional responses, thereby enhancing overall user satisfaction and reducing the cognitive load on users \cite{li2024user}. By integrating these advanced matching techniques, designers can create more fluid and adaptive AI-driven UIs that anticipate and react to user needs more effectively.

\section{Multimodal Interaction}

Generative AI is allowing for development of new UIs, bridging multiple input types to create more natural, contextually adaptive interactions. However, the integration of these systems presents unique challenges, especially with maintaining privacy, ensuring efficient processing on resource-limited mobile hardware, and retaining context across diverse inputs. Multimodal LLMs need optimized, lightweight frameworks to enable fluid, real-time interaction, demanding innovative solutions that balance cloud and edge processing.

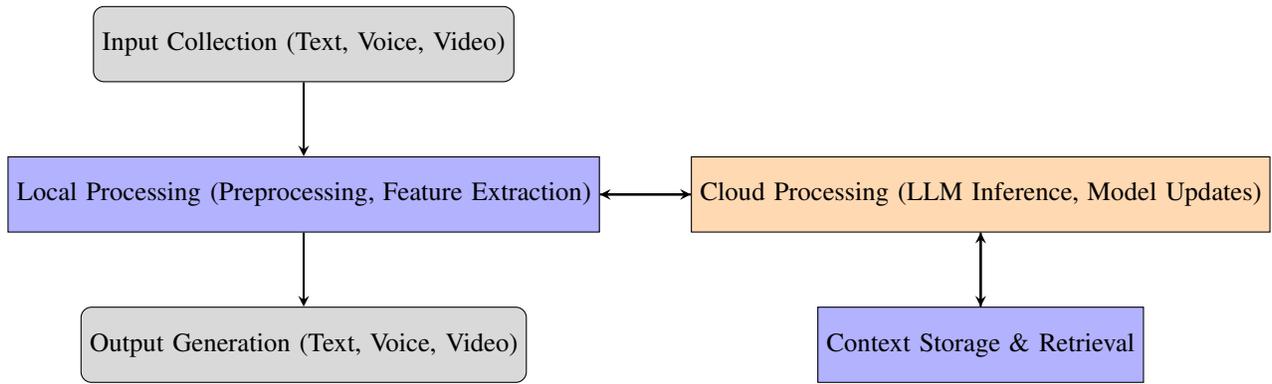
\begin{figure*}[h]
\centering
\begin{tikzpicture}[node distance=2cm]

\node (input) [startstop] {Input Collection (Text, Voice, Video)};
\node (localproc) [process, below of=input] {Local Processing (Preprocessing, Feature Extraction)};
\node (cloudproc) [cloud, right of=localproc, xshift=7cm] {Cloud Processing (LLM Inference, Model Updates)};
\node (output) [startstop, below of=localproc] {Output Generation (Text, Voice, Video)};
\node (storage) [process, below of=cloudproc] {Context Storage \& Retrieval};

\draw [arrow] (input) -- (localproc);
\draw [arrow] (localproc) -- (output);
\draw [arrow] (localproc) -- (cloudproc);
\draw [arrow] (cloudproc) -- (storage);
\draw [arrow] (storage) -- (cloudproc);
\draw [arrow] (cloudproc) -- (localproc);

\end{tikzpicture}
\caption{System architecture for lightweight framework on mobile devices.}
\label{fig:lightweight_architecture}
\end{figure*}

\subsection{Human-Computer Interface for Multimodal LLMs}

As Multimodal LLMs evolve, they increasingly require optimized interfaces to facilitate user interaction. The interface problem presents a critical question: what form should this interaction take? It seems that the optimal interface depends on the target user as well as the complexity of the task involved.
\begin{itemize}
    \item Console-based interfaces are ideal for technical users who prioritize control and efficiency, offering precise command-line input. However, they are less accessible to general users, require more knowledge to use them, and do not take advantage of AI capabilities.
    \item GUIs provide a more intuitive and user-friendly experience, especially for diverse user bases. However, they may be limited in multimodal settings requiring fluid processing of various input types. Songqin et al. propose a large multimodal model for mobile GUIs that emphasizes privacy-preserving data handling and the use of hybrid visual encoders, which adapt to different resolutions and layouts. This approach allows for nuanced interactions across diverse GUIs without sacrificing privacy or accuracy in handling multilingual content \cite{nong2024mobileflowmultimodalllmmobile}.
    \item Voice-based interfaces offer significant flexibility and can enhance user experience through natural, conversational exchanges. However, these systems require optimal acoustic conditions and often struggle with complex, high-dimensional tasks. As Badr et al. highlight, challenges arise in real environments, where voice-based systems must contend with background noise, diverse speaker characteristics, and varying signal conditions \cite{badr2020review}.
    \item Immersive interfaces (VR/AR) offer a highly interactive experience, but their accessibility, cost, and comfort limitations restrict their use to specialized contexts. Niu et al. present a great overview of how multimodal systems can improve intuitiveness and enhance user experience in Computer-Aided Design (CAD) \cite{app12136510}.
    \item Smart Space-based interfaces: Smart spaces, equipped with various sensors (e.g., cameras), offer users the ability to interact in an intuitive, gesture-based manner without the need for physical controllers \cite{lyu2024smart}. Such environments support natural user interactions by capturing gestures and contextual cues through embedded sensors, enabling seamless communication between users and the system \cite{lee2006universal}. This approach aligns with multimodal interaction goals by enabling hands-free control, especially beneficial in contexts where direct device handling may be challenging or unnecessary.
    \item Gesture-based interaction: Beyond fully equipped smart spaces, gesture-based interactions in conventional environments (e.g., smart check-out systems) utilize camera-based systems to interpret gestures, allowing users to interact with digital interfaces naturally. Such setups enable touch-free control without relying on substantial physical infrastructure \cite{wu2016intelligent}, presenting an accessible, cost-effective option for gesture-based interaction.

\end{itemize}

\subsection{Hardware Considerations: Mobile Phones}

The deployment of LLMs on mobile devices brings unique hardware challenges due to limitations in processing power, memory, and energy efficiency. Maintaining state-of-the-art functionality without compromising performance is crucial, as mobile phones are the primary devices for human-AI interaction. A key solution is the integration of efficient memory management techniques, like KV cache compression and chunk-level memory optimization, which can significantly reduce latency in LLM processing by managing memory at a granular level. Such innovations enable smoother, stateful LLM services on mobile devices, minimizing switching delays and enhancing user experience without exhausting memory resources \cite{yin2024llmservicemobiledevices}. 

One of the primary constraints in mobile LLM implementation is the device's limited energy and processing capacity, which necessitates lightweight model architectures and the use of accelerators like Neural Processing Units (NPUs). NPUs have been instrumental, achieving up to a 22× speedup over CPUs, thus facilitating real-time AI functionalities even on constrained devices \cite{yuan2023mobile}. Advances in model quantization, mainly down to 8- or even 4-bit precision, are further optimizing these models, allowing robust LLMs to operate with minimal memory and processing demands. For instance, a 3-billion-parameter GPT model has demonstrated efficient performance on devices with just 4GB of RAM, exemplifying how quantization can enable powerful on-device AI within tight resource limits \cite{carreira2023revolutionizingmobileinteractionenabling}.

To help examine the performance of different LLM-based mobile agents, Deng et al. \cite{deng2024mobilebenchevaluationbenchmarkllmbased} propose a benchmarking framework that provides realistic test scenarios by integrating both UI operations and API calls, ensuring comprehensive evaluation of these models in practical, user-centered tasks. Such benchmarks highlight the importance of balancing computational efficiency with user demands and allow for finding areas that require further optimization. Further, the proposed approach of treating mobile LLMs as OS-level services further centralizes AI functionalities, promoting better resource sharing and reducing redundancy, as multiple applications can access a single LLM instance rather than each maintaining separate models \cite{chen2024llmmobileinitialroadmap}.

Lastly, smart spaces provide a distinct advantage in multimodal interaction contexts by leveraging distributed sensor infrastructure to reduce processing demands on mobile devices \cite{chio2023smartspec}. Through ambient sensing, these environments capture user input, interpret gestures, and offer responsive interactions with minimal reliance on local device resources \cite{lee2006universal}. Gesture-based systems, like intelligent self-checkout setups, similarly reduce hardware constraints by centralizing processing in external infrastructure \cite{wu2016intelligent}. This distribution of resources can significantly alleviate the challenges imposed by mobile hardware limitations, especially for applications requiring continuous interaction.

\subsection{Lightweight Frameworks for Multimodal Systems}

A lightweight framework is essential for integrating AI-driven, multimodal systems into mobile applications. This section will explore potential frameworks that are scalable and efficient, minimizing computational overhead while maximizing the multimodal capabilities of the system.

Figure \ref{fig:lightweight_architecture} illustrates the system architecture for a lightweight framework designed for mobile devices. It demonstrates how inputs, such as text, voice, or video, are first processed locally through preprocessing and feature extraction. To optimize performance, more complex tasks, like LLM inference and model updates, are handled via cloud processing. The architecture also includes context storage and retrieval mechanisms in the cloud to ensure that interactions remain contextually coherent over time. This architecture effectively balances local and cloud processing, making it suitable for resource-constrained devices while ensuring responsive, multimodal AI-driven interactions.

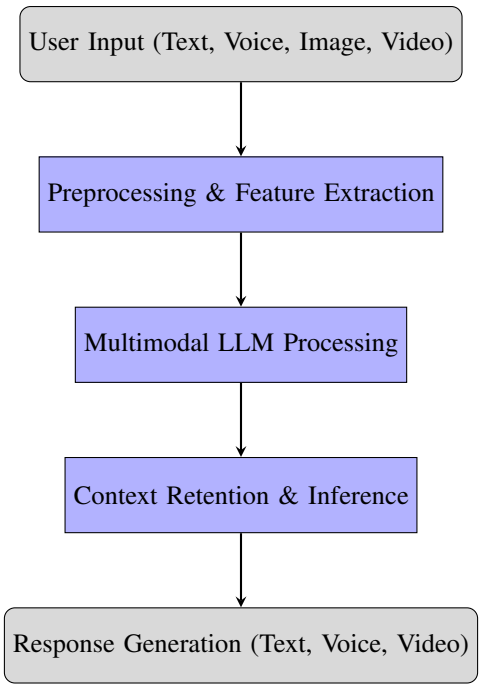
\begin{figure}[h]
\centering
\begin{tikzpicture}[node distance=2cm]

\node (input) [startstop] {User Input (Text, Voice, Image, Video)};
\node (preprocess) [process, below of=input] {Preprocessing \& Feature Extraction};
\node (multimodal) [process, below of=preprocess] {Multimodal LLM Processing};
\node (context) [process, below of=multimodal] {Context Retention \& Inference};
\node (output) [startstop, below of=context] {Response Generation (Text, Voice, Video)};

\draw [arrow] (input) -- (preprocess);
\draw [arrow] (preprocess) -- (multimodal);
\draw [arrow] (multimodal) -- (context);
\draw [arrow] (context) -- (output);

\end{tikzpicture}
\caption{Workflow of AI-driven multimodal interaction.}
\label{fig:ai_multimodal_workflow}
\end{figure}

Figure \ref{fig:ai_multimodal_workflow} illustrates the workflow of AI-driven multimodal interaction. User inputs in the form of text, voice, image, or video undergo preprocessing and feature extraction to facilitate efficient processing by the multimodal LLM. The system then applies context retention and inference mechanisms to maintain continuity across interactions, before generating an appropriate response, which may include text, voice, or video outputs. This structured approach ensures that the system can handle diverse inputs while preserving context and delivering relevant responses.


\section{Limitations, Challenges, and Future Directions for AI-driven Interfaces}

The rapid advancements in generative AI and multimodal interfaces present many opportunities for enhancing user experiences and interaction technologies. As these innovations continue to evolve, the future of UIs will be shaped by greater automation, dynamic adaptability, and the incorporation of novel input modalities. In this section, we explore the current limitations, potential directions, and solutions that leverage these emerging technologies to create more intuitive, efficient, and contextually aware UIs, paving the way for transformative user experiences across various platforms and devices.

\subsection{Current Limitations and Challenges}

\subsubsection{Technical constraints}

Implementing AI-driven UIs faces significant technical challenges, especially regarding real-time performance and memory retention. One primary limitation is achieving low-latency responses while handling multiple, concurrent input types, such as text, voice, and video, seamlessly within multimodal interfaces. Real-time performance is critical to maintaining user engagement and satisfaction, yet processing multimodal inputs at such speeds often requires substantial computational resources, typically available only on high-end hardware.

A further challenge is designing lightweight frameworks that can operate effectively on mobile devices, which are typically constrained by limited processing power, memory, and battery life. Although cloud-based processing can alleviate some of these demands, it introduces latency and potential privacy concerns. Thus, developing on-device AI capabilities that retain multimodal LLMs functionality without sacrificing performance remains an area for ongoing research. Lightweight frameworks optimized for mobile devices must balance computational efficiency with the robust features necessary to process diverse input types in real-time. Techniques such as model quantization, memory optimization, and parameter-efficient tuning are essential to supporting the full capabilities of multimodal LLMs on mobile platforms.

\subsubsection{Ethical considerations}

AI-driven interfaces also present ethical challenges, particularly around data privacy, transparency, and user trust. As AI systems collect and process vast amounts of personal data to adaptively enhance user experiences, they risk compromising user privacy. This challenge is compounded in multimodal interfaces, where various data types—such as voice, text, and video—are processed simultaneously, increasing the volume and sensitivity of collected information.

One major ethical concern is the ``black-box" nature of many AI models, which often operate without clear insight into how they reach certain decisions. This lack of transparency can reduce user trust and create barriers to adoption, especially in sensitive domains like healthcare or finance. To address this, explainable AI (XAI) techniques are essential, providing users with understandable insights into AI decision-making processes. Additionally, the risk of bias in AI responses is another ethical issue. Without careful design and training on diverse datasets, AI-driven interfaces may exhibit biased behavior, potentially leading to discriminatory outcomes. Incorporating fairness algorithms and extensive dataset diversity during training can mitigate these biases.

For mobile deployments, resource limitations complicate these ethical considerations further. Processing sensitive data on mobile devices, for instance, requires privacy-preserving methods, such as data anonymization and secure storage. Table \ref{tab:ethical_technical_challenges} outlines key ethical and technical challenges in AI-driven UIs, offering solutions like data encryption, anonymization, and transparent user control mechanisms. Addressing these concerns is crucial not only for protecting user privacy but also for building trust in AI-driven systems, especially as they become more integrated into everyday mobile applications.

\subsection{Emerging Trends and Future Directions}

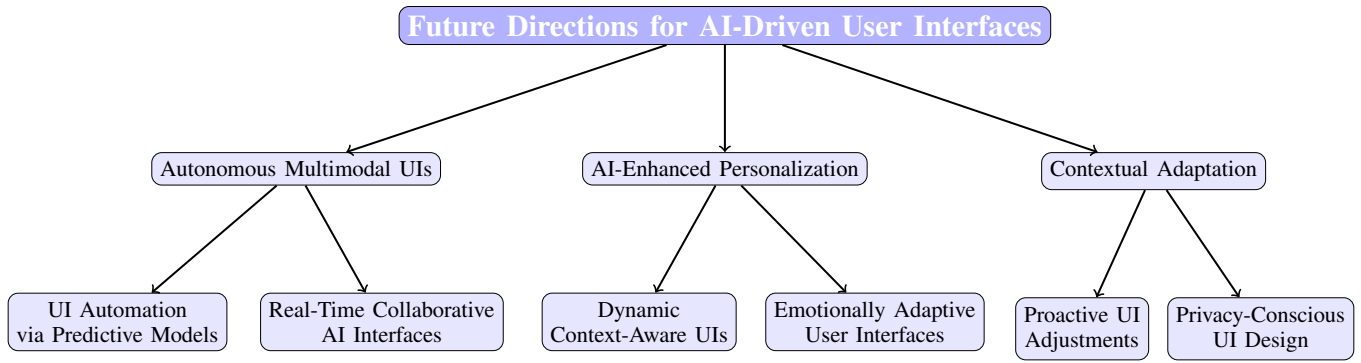
\begin{figure*}[h!]
    \centering
    \resizebox{\textwidth}{!}{
    \begin{tikzpicture}[
        node distance=1.5cm and 2cm,
        every node/.style={rectangle, rounded corners, draw=black, align=center, font=\small, fill=blue!10},
        every label/.style={font=\small\bfseries},
        level 1/.style={sibling distance=7cm},
        level 2/.style={sibling distance=3cm},
        edge from parent/.style={draw, -latex, thick},
        ]

        \node[fill=blue!30, text=white, font=\large\bfseries] (root) {Future Directions for AI-Driven User Interfaces};

        \node [below=of root, xshift=-6cm] (autonomous) {Autonomous Multimodal UIs};
        \node [below=of root, xshift=0cm] (personalization) {AI-Enhanced Personalization};
        \node [below=of root, xshift=6cm] (contextual) {Contextual Adaptation};

        \node [below=of autonomous, xshift=-2.5cm] (automation) {UI Automation \\ via Predictive Models};
        \node [below=of autonomous, xshift=1.2cm] (collaboration) {Real-Time Collaborative \\ AI Interfaces};

        \node [below=of personalization, xshift=-1.2cm] (dynamic_ui) {Dynamic \\ Context-Aware UIs};
        \node [below=of personalization, xshift=2.1cm] (emotion) {Emotionally Adaptive \\ User Interfaces};

        \node [below=of contextual, xshift=-1cm] (proactive) {Proactive UI \\ Adjustments};
        \node [below=of contextual, xshift=1.5cm] (privacy) {Privacy-Conscious \\ UI Design};

        \draw[->, thick] (root) -- (autonomous);
        \draw[->, thick] (root) -- (personalization);
        \draw[->, thick] (root) -- (contextual);

        \draw[->, thick] (autonomous) -- (automation);
        \draw[->, thick] (autonomous) -- (collaboration);

        \draw[->, thick] (personalization) -- (dynamic_ui);
        \draw[->, thick] (personalization) -- (emotion);

        \draw[->, thick] (contextual) -- (proactive);
        \draw[->, thick] (contextual) -- (privacy);

    \end{tikzpicture}
    }
    \caption{Taxonomy of future directions in AI-driven UIs.}
    \label{fig:taxonomy}
\end{figure*}

Figure \ref{fig:taxonomy} illustrates a taxonomy of future directions in AI-driven UIs, categorizing potential advancements such as dynamic, context-aware UIs and emotionally adaptive interfaces. Detailed discussion on these future directions are provided next.

\subsubsection{AI-automated interfaces and dynamic contextual UIs}

The future of UIs is moving towards full automation, where UIs can dynamically adapt to user behavior and environmental context without requiring manual customization. AI-automated interfaces leverage real-time data on user preferences, actions, and situational factors—such as location, activity, or emotional state—to adjust elements like layout, functionality, and interaction methods accordingly. By continuously learning from user interactions, these UIs can provide seamless and anticipatory responses, enhancing accessibility and usability while minimizing cognitive load.

Dynamic, context-aware UIs represent a significant step forward in creating personalized and responsive interfaces. These systems can modify interface elements based on live feedback, adjusting to changes in lighting, movement, or even user mental states. Lightweight frameworks and advancements in edge computing and federated learning play a crucial role in enabling these adaptive interfaces to function efficiently on mobile devices, which are often constrained by limited processing capabilities. By processing sensitive data locally on the device, these frameworks preserve user privacy, ensuring that dynamic, contextually adaptive UIs can provide real-time personalization without compromising security or performance.

\subsubsection{New modalities and hybrid approaches}

Emerging interaction modalities are paving the way for innovative and immersive experiences, with technologies such as brain-computer interfaces (BCIs), gesture recognition, and haptic feedback leading the charge. BCIs offer a groundbreaking approach to interaction by potentially enabling users to control digital systems through neural input, facilitating hands-free and thought-based control. While still in developmental stages, BCIs have promising applications in accessibility tools for users with mobility impairments, as well as in immersive environments like gaming and virtual reality.

Gesture recognition and haptic feedback are also gaining traction, particularly in AR and VR applications. Gesture recognition allows for intuitive, hands-free interaction by interpreting physical movements, while haptic feedback adds a tactile dimension, enhancing the immersive quality of user experiences. Together, these modalities create a more engaging and accessible interface, catering to a wide range of user preferences and needs.

Hybrid interfaces that combine traditional GUIs with multimodal inputs offer a versatile solution to the challenge of context retention and seamless transitions across input types. These hybrid UIs allow users to switch between text, voice, and visual interactions, creating a flexible and fluid interaction environment, especially on mobile devices where screen space is limited. Leveraging multimodal fusion techniques, these interfaces can integrate and align diverse input types—text, voice, images—in a cohesive manner. This integration not only enriches the interaction experience but also allows for more effective handling of complex tasks, as the system can adapt to user preferences and contextual cues dynamically.

Overall, these emerging modalities and hybrid approaches reflect a trend toward more adaptive, intuitive, and accessible UIs, driven by AI advancements that prioritize real-time responsiveness and a seamless user experience across platforms.

\subsection{Proposed Solutions and Innovations}

\subsubsection{AI-generated UIs and proactive adjustments}

Generative AI offers a transformative potential to create UIs that are dynamically tailored to real-time user behavior and contextual cues, leading to more personalized and efficient interaction spaces. By analyzing user habits, preferences, and situational factors, generative AI models can adapt UIs to prioritize relevant information, simplify navigation, and reduce cognitive load. This enables systems to provide user-specific experiences that evolve based on interaction patterns, ultimately enhancing accessibility and usability.

Predictive modeling is another key innovation driving proactive UI adjustments. Leveraging past user interactions, these models can anticipate user needs, making adjustments to interface components such as shortcuts, layout, and displayed content before the user initiates a task. This proactive approach streamlines workflows, minimizes navigation time, and provides a more intuitive interaction experience, especially valuable in professional environments like design, healthcare, and engineering, where efficiency is critical.

\subsubsection{Emotionally adaptive UIs and autonomous collaboration}

Emotionally adaptive UIs represent a frontier in responsive design, enabling interfaces to adjust based on real-time assessments of user emotional states. By analyzing multimodal inputs—such as facial expressions, tone of voice, and physiological cues—these UIs can modify design elements (e.g., color scheme, layout) and functionality to create a more supportive and comfortable experience. This capability is particularly beneficial in applications like mental health support and customer service, where responding to user emotions can significantly improve engagement and satisfaction.

In complex, collaborative settings, such as healthcare operating rooms, autonomous multimodal collaboration plays an essential role. These AI-driven UIs can seamlessly integrate multiple input types—text, voice, gesture, and visual cues—allowing hands-free control and intuitive interaction in high-stakes environments. Autonomous multimodal collaboration facilitates the efficient exchange of information, improves workflow coordination, and reduces the cognitive burden on users, enabling smoother and more effective teamwork.

\subsubsection{Ethical, privacy-conscious design, and cross-platform AR UIs}

As AI-driven UIs become more pervasive, designing ethically responsible interfaces is essential. Privacy-conscious design frameworks prioritize user autonomy by providing real-time transparency and control over data usage. AI systems embedded within these UIs can implement data anonymization techniques, generate privacy-preserving interaction logs, and give users greater agency over their personal information, fostering trust and transparency.

Cross-platform AR UIs are another innovative solution, creating a seamless and consistent experience across devices by integrating spatial data to bridge digital and physical environments. These AR interfaces enable users to interact with virtual elements that appear in their physical surroundings, offering continuity whether accessed via mobile devices, AR glasses, or desktop systems. This approach enhances accessibility and user experience, making AR tools more adaptable and practical for everyday applications.

\subsubsection{Collaborative AI for real-time co-creation}

Collaborative AI introduces the possibility of real-time co-creation, where users and AI work together in a highly interactive, immersive interface. In creative fields such as design, video editing, and educational content creation, these collaborative UIs allow users to input ideas across multiple modalities (e.g., text, speech, sketching) while AI autonomously augments and builds upon these inputs. The AI can provide real-time suggestions, design enhancements, or generate elements that complement the user's contributions, resulting in an interactive co-creation process.

This real-time collaboration with AI expands user capabilities and shortens design cycles, allowing for a more engaging, productive, and immersive experience. Collaborative UIs can act as intelligent partners, assisting users in generating high-quality content quickly while adapting to their creative style and preferences. This co-creation capability highlights the potential for AI to empower users in producing complex, innovative solutions, fostering creativity, and enhancing educational tools through interactive learning experiences.








\begin{table*}[h]
\centering
\caption{Ethical and technical challenges in AI-driven UIs.}
\begin{tabularx}{\textwidth}{|l|X|X|X|X|}
\hline
\textbf{Challenge} & \textbf{Description} & \textbf{Ethical Concern} & \textbf{Technical Limitation} & \textbf{Possible Solutions} \\ \hline
\textbf{Privacy} & Collection of personal data through interactions & User data misuse & Data storage \& encryption & Data anonymization, secure storage \\ \hline
\textbf{Transparency} & Lack of clarity in AI decision-making & AI black-box problem & Complex LLM architectures & Explainable AI, model interpretability \\ \hline
\textbf{Bias in AI} & Discriminatory behavior in AI responses & Amplification of social biases & Biased training data & Fairness algorithms, diverse datasets \\ \hline
\textbf{User Trust} & Trust issues due to unpredictable AI behavior & Loss of user confidence & Lack of context retention & Transparent decision-making, user control \\ \hline
\textbf{Processing Power} & High computational requirements for multimodal LLMs & Energy consumption, hardware limitations & Limited processing power on mobile devices & Lightweight models, edge computing \\ \hline
\end{tabularx}
\label{tab:ethical_technical_challenges}
\end{table*}


\section{Metrics for Evaluating AI-Driven Multimodal UIs}

Evaluating the effectiveness of AI-driven, multimodal UIs requires standardized metrics to assess performance across different dimensions, including response accuracy, latency, user retention, and feedback quality. By implementing these metrics, researchers and developers can ensure consistent, reliable evaluations across various applications and platforms.

\subsection{Accuracy of Response in Different Modalities}

A critical metric for AI-driven, multimodal UIs is the accuracy of responses across different input types (e.g., text, voice, image). This includes evaluating how well the system interprets and processes user inputs in each modality. For instance, voice input accuracy is often measured through word error rate (WER), while image recognition accuracy may use metrics like precision, recall, and F1-score to evaluate the system's ability to interpret visual data accurately \cite{wang2024applicationartificialintelligencehand, li2024user}.

\subsection{Latency}

Latency is another essential metric, reflecting the response time of the UI from input recognition to output generation. Low latency is critical for seamless interactions, particularly in real-time applications like VR or AR. Measuring latency involves calculating the average response time across different tasks, with lower values indicating better performance. Studies suggest that latency under 100ms is generally acceptable for multimodal UIs, especially in applications involving voice or touch-based inputs \cite{Bandi2023, carreira2023revolutionizingmobileinteractionenabling}.

\subsection{User Retention}

User retention is a qualitative metric that assesses how well the interface engages users over time. High user retention indicates that the UI is effective, user-friendly, and encourages users to return. This metric can be measured by tracking the frequency and duration of interactions within the system. For AI-driven, multimodal UIs, retention may vary by modality, as some users may prefer one interaction method over another \cite{huang2024unlocking, zhang2024unveiling}.

\subsection{Feedback Quality}

Qquality measures the system's ability to incorporate user feedback effectively to improve future interactions. For multimodal UIs, feedback may include explicit user ratings or implicit signals, such as abandonment of certain interaction paths. Systems that can adapt based on user feedback are considered more effective, as they evolve with user preferences, contributing to overall satisfaction and engagement \cite{Su2023, kim2021multimodal}.

\subsection{Suggested Evaluation Methodologies}

To standardize assessments of AI-driven, multimodal UIs across applications, the following evaluation methodologies are recommended.

\subsubsection{A/B testing across modalities} 

A/B testing allows researchers to compare performance variations in different interaction modalities (e.g., text vs. voice). By tracking metrics like accuracy and latency across these versions, A/B testing can reveal the most effective modality for specific tasks \cite{li2024map, torricelli2024role}.

\subsubsection{Task-specific benchmarking} 

Benchmarking against established tasks allows researchers to assess the UI's performance in specific scenarios, such as conversational AI or AR interactions. Task-specific benchmarking provides a structured approach to evaluate metrics like latency and feedback quality, highlighting areas for improvement \cite{nong2024mobileflowmultimodalllmmobile, groner2024investigating}.

\subsubsection{User experience surveys} 

Collecting qualitative data through user experience surveys can help evaluate subjective metrics, such as user retention and feedback quality. Surveys provide insights into user satisfaction, preferred modalities, and perceived ease of use, which are essential for refining AI-driven UIs \cite{hartson1989human, groner2024investigating}.

\subsubsection{Longitudinal studies for retention analysis} 

Longitudinal studies, where user interactions are monitored over time, are crucial for understanding retention dynamics and engagement patterns. By tracking metrics like user retention rates over weeks or months, these studies offer insights into long-term UI effectiveness \cite{liu2024mental, mridha2021brain}.

By applying these metrics and methodologies, developers and researchers can better assess the performance of AI-driven, multimodal UIs, facilitating continuous improvement and enabling consistent comparisons across different systems and applications.

\section{Conclusion}

The integration of Generative AI into modern user interfaces marks a transformative shift in human-computer interaction, creating new pathways for personalized, multimodal, and cross-platform experiences. This review has highlighted the evolution of UIs from traditional single-modal interfaces to adaptive, multimodal systems powered by large language models (LLMs) and other generative technologies. By examining the interface dilemma—balancing text, voice, visual, and immersive interactions within a single framework—this paper sheds light on the critical challenges and design considerations essential for effective multimodal UIs.

Moreover, the need for lightweight, mobile-compatible frameworks emphasizes the importance of scalable solutions that operate seamlessly across devices. Such frameworks must address not only technical challenges, such as latency and limited processing power, but also ethical issues related to privacy, context retention, and responsible data usage. These challenges highlight the need for careful design in managing both cloud-based and on-device AI resources to ensure privacy, efficiency, and user trust in AI-powered systems.

Looking forward, Generative AI holds immense potential for advancing user-centric and emotionally adaptive interfaces that respond to real-time cues, user preferences, and predictive modeling. Future research should focus on refining multimodal interfaces to accommodate diverse user needs while ensuring ethical safeguards and computational efficiency. As Generative AI continues to evolve, it is positioned to redefine the boundaries of interactive experiences, setting a new standard for adaptive, accessible, and intelligent UIs across digital platforms.


\bibliographystyle{IEEEtran}
\bibliography{refs}


\end{document}